\documentclass{aa}
\usepackage{graphics}
\usepackage{times}
\begin{document}
\thesaurus{12(12.04.3; 11.04.1;
11.07.1; 11.11.1; 11.12.1)}
\title{
On the quiescence of the Hubble flow in the vicinity of \\ the Local Group
}
\subtitle{A study using galaxies with distances from the Cepheid
PL-relation
}
\author{
T.~Ekholm\inst{1,3}
\and
Yu.~Baryshev\inst{2}
\and
P.~Teerikorpi\inst{3}
\and
M.~O.~Hanski\inst{3}
\and
G.~Paturel\inst{1}
}
\offprints{
T.~Ekholm
}
\institute{
CRAL - Observatoire de Lyon,
F69561 Saint Genis Laval CEDEX, France
\and
Astronomical Institute, St.Petersburg State University, Staryj Petergoff,
198504, St.Petersburg, Russia
\and
Tuorla Observatory,
FIN-21500 Piikki\"o,
Finland
}
\date{received, accepted}
\maketitle
%
%
%
%
\begin{abstract}
Cepheid distances of local galaxies ($<7\mathrm{\ Mpc}$) are used to study
the very nearby velocity field, as pioneered by
Sandage (\cite{Sandage86}) 
who also pointed out its remarkable properties:
linearity and quietness. The new data show that the velocity
dispersion in the distance range as seen from the barycentre
of the Local Group 
$1-8\mathrm{\ Mpc}$ is as low as $38\mathrm{\ km\, s^{-1}}$.
The local rate of expansion coincides with the global Hubble
constant. Down to 1.5 Mpc we cannot detect a deviation from the
linear Hubble flow. This puts an upper limit for the mass of the
Local Group, for a wide class of Friedman models, including
those with the cosmological constant.

\keywords{Cosmology: observations --
Galaxies: distances and redshifts -- Galaxies: general --
Galaxies: kinematics and dynamics -- Local Group}
\end{abstract}
%
%
%
%
\section{Introduction}

In his classical paper Sandage 
(\cite{Sandage86}) studied the perturbation of the
very nearby velocity field using the available distances to local
galaxies, mostly obtained by ground-based observations of brightest resolved
stars and Cepheids.  He concluded that the deviation from the linear
Hubble flow possibly was detected at very small distances, while the
quiet, linear Hubble law starts at about 2 Mpc. He pointed out
the remarkable properties of the local velocity field: 1) The linearity
of the velocity-distance relation down to small distances, 2) the closeness
of the local and global rates of expansion, and 3) the small velocity
dispersion around the Hubble law.  He also asked what happens when
the distances become more accurate, in particular, will the velocity
dispersion still decrease below 
$\sigma_v = 60\mathrm{\ km\, s^{-1}}$ which he derived
from those data. We also note that Karachentsev \& Makarov
(\cite{Kara96}) found that by using both massive and small galaxies
the velocity dispersion in the local volume is 
$\sigma_v = 72\mathrm{\ km\, s^{-1}}$.

The aim of our Letter is to use new accurate data on the distances of
nearby galaxies inferred for the extragalactic Cepheid $PL$-relation
(cf. Sect.~2) in order to study further the behaviour of the Hubble
law in the outskirts of the Local Group.
\section{The sample}
We make use of the the Lyon Extragalactic Cepheid Database
(Lanoix et al.
\cite{Lanoix99b}). In this database there are
32 galaxies with both V and I band photometry. 
23 galaxies have Cepheid
measurement from the HST and rest have groundbased measurements.
The $PL$ relation was calibrated using a sample of galactic
Cepheids measured with the astrometric HIPPARCOS satellite
(Lanoix et al. \cite{Lanoix99a}) and corrected for 
a Malmquist-like bias according
to Lanoix et al. (\cite{Lanoix99c}).

As we are interested in the very local velocity field
we restricted our sample by requiring
$R_\mathrm{gal}\le7\mathrm{\ Mpc}$, where also
the Virgo-centric flow correction is small.
$R_\mathrm{gal}$ refers to the distance in Mpc measured from
Galaxy. 
We were left with 14 galaxies. 
The Virgo flow at larger
distances was studied with Cepheid-based distances in Ekholm et al.
(\cite{Ekholm99a}).

In Table~1 we present the data relevant for the present analysis.
In column 1 we give the PGC number and
in column 2 we give the name. In column 3 we give the
distance in Mpc as calculated from the distance moduli provided in
the Lyon Extragalactic Cepheid Database. In column 4 we give the 
distance in Mpc as seen from the barycentre of the Local Group.
In columns 5-7 we give the velocities: 
$V_{\sun}$ is the mean heliocentric velocity
extracted from LEDA (the Lyon-Meudon Extragalactic Database), 
$V_\mathrm{Yahil}$ is $V_{\sun}$ corrected for the centroid of the
Local Group according to Yahil (\cite{Yahil77}) and 
$V_\mathrm{corr}$ is $V_\mathrm{Yahil}$
corrected for the virgocentric motion.
We explain these corrections in the next section.
As regards uncertainties in the parameters we note that the mean
error for $V_{\sun}$ is $6.4\mathrm{\ km\,s^{-1}}$. In Fig.~\ref{F1}
we plot the error bars reflecting the uncertainties in the
Cepheid distances. 

Finally we remind that \object{LMC} and
\object{NGC6822} belong to the Milky Way subsystem and
\object{NGC598} (\object{M33}) is a member of the \object{M31}
subsystem. The status of the irregular galaxy \object{IC1613}
is unclear.
Gurzadian et al. (\cite{Gurzadian93}) and  Rauzy \&
Gurzadian (\cite{Rauzy98}) link it with the Galaxy, but 
van den Bergh (\cite{vdBergh99}) recognizes that
\object{IC1613} may be a free-floating member
of the Local Group. Thus it is wise to consider it as a special
case in our diagrams.
%

\begin{table}
\caption{In this table we present the the crucial data for the
14 galaxies with extragalactic $PL$-distances. 
The entries are explained in the text.
}
\begin{center}
\begin{tabular}{llccccc}
\hline
LEDA & NAME  & $R_\mathrm{gal}$ & $R_\mathrm{2/3}$ 
& $V_{\sun}$ & $V_\mathrm{Yahil}$ & $V_\mathrm{corr}$\\
\hline
3844  & IC1613  & 0.69 & 0.44 & -231 & -62 & -72 \\
45314 & IC4182  & 4.94 & 5.09 &  316 & 337 & 274 \\
17223 & LMC     & 0.05 & 0.61 &  317 &  81 &  82 \\
2557  & N224    & 0.87 & 0.29 & -300 & -13 & -12.7 \\
3238  & N300    & 2.17 & 2.13 &  142 & 125 & 101 \\
28630 & N3031   & 3.37 & 3.18 &  -35 & 125 & 147 \\ 
29128 & N3109   & 1.02 & 1.53 &  404 & 130 & 131 \\
34554 & N3621   & 6.61 & 7.16 &  727 & 436 & 457 \\
48334 & N5253   & 3.16 & 3.73 &  403 & 155 & 153 \\
50063 & N5457   & 6.92 & 6.87 &  240 & 361 & 374 \\ 
5818  & N598    & 0.79 & 0.27 & -180 &  69 &  66 \\
63616 & N6822   & 0.45 & 0.73 &  -56 &   8 &  11 \\
29653 & SEXA    & 1.45 & 1.88 &  325 & 118 & 110 \\
28913 & SEXB    & 1.39 & 1.76 &  302 & 139 & 131 \\
\hline
\end{tabular}
\end{center}
\end{table}
\section{The method}
To study the local Hubble flow at smaller distances, it becomes increasingly
important to make relevant corrections to
distances and radial velocities. The first correction, discussed by
Sandage (1986), is due to the shift
of the observer to the centre of expansion, which in the
self-gravitating Local Group is not in our Galaxy, but presumably in
the barycentre.
 By assuming the mass ratio of \object{M31} and
\object{Galaxy} is $M_{M31}/M_{G}\approx2$ Sandage set the
barycentre to be on the line between \object{M31} and
\object{Galaxy} at 2/3 of the distance to \object{M31}.
This is also our choice. We denote the distance from the
barycentre by $R_{2/3}$.

The observed mean heliocentric velocities must also be corrected.
We first correct the
observed velocity to the value as it would be measured by an 
observer in our Galaxy
being at rest relative to the centroid of the Local Group\footnote{
We note that the observer at the centroid would measure a slightly
different velocity, depending on the 
location of the galaxy and the nature of the velocity 
field close to the Local Group. For example, if the deviation of
the velocity field from the Hubble flow is exactly spherically symmetric
relative to the centroid, one may show by starting from the first
principles of the Hubble velocity field that then the observer at the
centroid would see the velocity $V_\mathrm{Yahil}/\cos\theta$. $\theta$
is the angle between the centroid and the Galaxy as seen 
from the measured galaxy. In the present case, $\theta$ is normally
quite small. Hence such a correction makes the velocity usually only
$1$-$5\mathrm{\ km\, s^{-1}}$ larger for our field galaxies. We prefer
here to use the simpler velocity correction as done by Sandage
(\cite{Sandage86}).}. 
This we
do according
to Yahil et al. (\cite{Yahil77}), which is also the preferred
choice in the LEDA database. We also remind that in Ekholm et al.
(\cite{Ekholm99a}) the correction was also made according to
Richter et al. (\cite{Richter87}) but difference to 
the correction of Yahil et al. (\cite{Yahil77}) was quite small. 
The velocity correction used in
reads:
%
\begin{equation}
\label{yts}
\Delta v = 295.4\sin l \cos b - 79.1\cos l \cos b - 37.6\sin b.
\end{equation}

After the solar motion is thus corrected for, the infall to the Virgo
Supercluster is the next disturbance. We correct the radial velocity
 for the virgocentric
motion following Ekholm et al. (\cite{Ekholm99a}):
%
\begin{eqnarray}
\label{vcor}
V_\mathrm{corr}=V_\mathrm{Yahil}&\pm&[v(d)_\mathrm{H}-v(d)]
\times \nonumber \\
 & & \sqrt{1-\sin^2\Theta/d^2}+V_\mathrm{LG}^\mathrm{in}\cos\Theta,
\end{eqnarray}
where ($-$) is valid for points closer than the tangential point
$d_\mathrm{gal}<\cos\Theta$ and ($+$) for
$d_\mathrm{gal}\ge\cos\Theta$.

$\Theta$ is the angular distance of a galaxy from the centre of Virgo
and $d$ is its distance $R$ from the centre normalized to the distance
of Virgo ($d=R/R_\mathrm{Virgo}$). Following
Ekholm et al. (\cite{Ekholm99a}) we take
$R_\mathrm{Virgo}=21\mathrm{\ Mpc}$  and
 $H_0=57\mathrm{\ km\,s^{-1}\,Mpc^{-1}}$. The Hubble velocity at a distance
$d$ from the centre  of Virgo is
$v(d)_\mathrm{H}=V_\mathrm{cosm}(1)\times d$.
The cosmological velocity
of the Virgo cluster becomes
$V_\mathrm{cosm}(1)=1200\mathrm{\ km\,s^{-1}}$.
The predicted velocity $v(d)$ is solved using the Tolman-Bondi model
as described by Ekholm et al. (\cite{Ekholm99a}).
\section{Results}
\begin{figure}
\resizebox{\hsize}{!}{\includegraphics{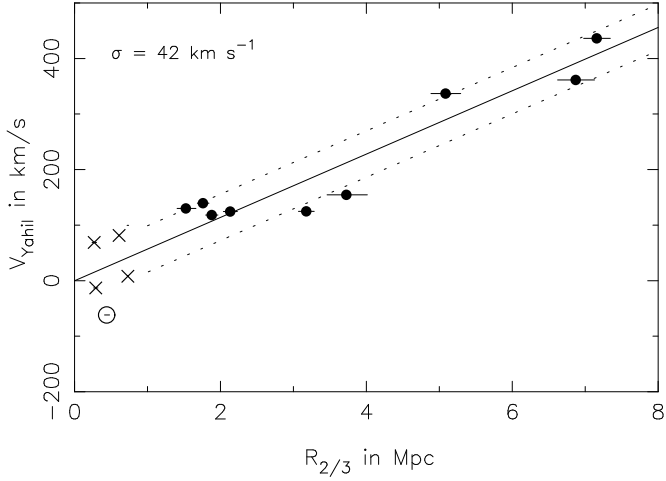}}
\caption{Velocities corrected to 
the rest frame of the centroid of the Local Group
according to Yahil et al. (\cite{Yahil77}). The origin of the
distances has been changed to the centroid following
Sandage (\cite{Sandage86}). 
The solid straight line is the
Hubble law for $H_0=57\mathrm{\ km\,s^{-1}\,Mpc^{-1}}$. The
dotted lines give the $1\sigma$ dispersion calculated for
galaxies between $1\mathrm{\ Mpc}$ and $8\mathrm{\ Mpc}$
giving $\sigma = 42\mathrm{\ km\,s^{-1}}$ Crosses refer to 
\object{LMC}, \object{N6822}
\object{N224} and \object{N598}. The open circle is \object{IC1613}.
The error bars reflect the formal uncertainties in the Cepheid
distances.
}
\label{F1}
\end{figure}

\begin{figure}
\resizebox{\hsize}{!}{\includegraphics{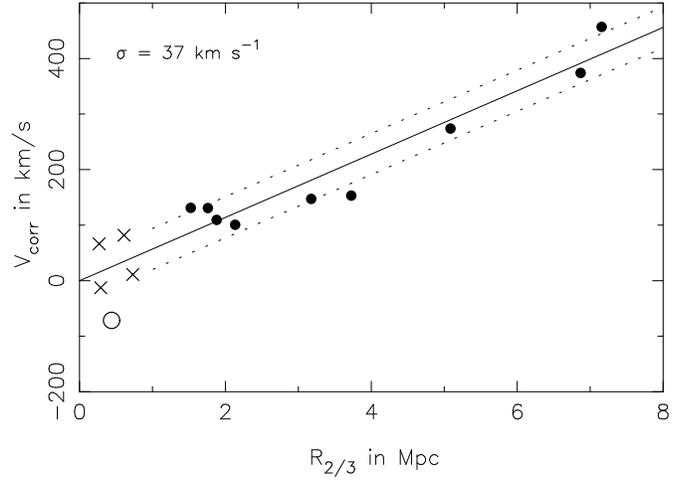}}
\caption{As Fig.~\ref{F1} except that now we have also corrected
for the Virgocentric motions. The effect is small except for
\object{IC4182}. This is the only galaxy that is at a small
angular distance from Virgo cluster. The velocity dispersion
$\sigma_v=37\mathrm{\ km\,s^{-1}}$.  
}
\label{F2}
\end{figure}

We present the corrected velocity vs. distance relations for the very
local velocity field ($R_{2/3} < 8$ Mpc) in Figs.~\ref{F1} and~\ref{F2}.
In Fig.~\ref{F1} velocities are corrected only for the solar motion
according to Yahil et al. (\cite{Yahil77}).  
We calculate the velocity dispersion  without
the Local Group members (the first 4 symbols marked as crosses), 
where certainly the
internal dynamics masks the possible Hubble term which is small compared
with the value of the velocity dispersion.
The velocity dispersion is
$\sigma_v=42\mathrm{\ km\, s^{-1}}$ solved from galaxies between
$R_{2/3}\in[1.0,8.0]\mathrm{\ Mpc}$. The expected Hubble law is given
as a solid straight line and the $1\sigma$ dispersion as dotted lines.

In Fig.~\ref{F2} we show the influence of the correction
for the virgocentric
motion. Now the velocity dispersion
decreases down to $\sigma_v=37\mathrm{\ km\,s^{-1}}$. 
The decrease in $\sigma_v$ is not significant. This is expected
because most of the galaxies are at large angular distance
from the Virgo except \object{IC4182} having $\Theta=26.4\degr$.
It is quite remarkable that after the correction this galaxy follows
almost exactly the expected Hubble law. 

Both Figs.~\ref{F1} and~\ref{F2} support the conclusion 4 of
Sandage (\cite{Sandage86}). The observed random velocities get smaller
as the uncertainties in the adopted distance indicator diminish.  

\begin{figure}
\resizebox{\hsize}{!}{\includegraphics{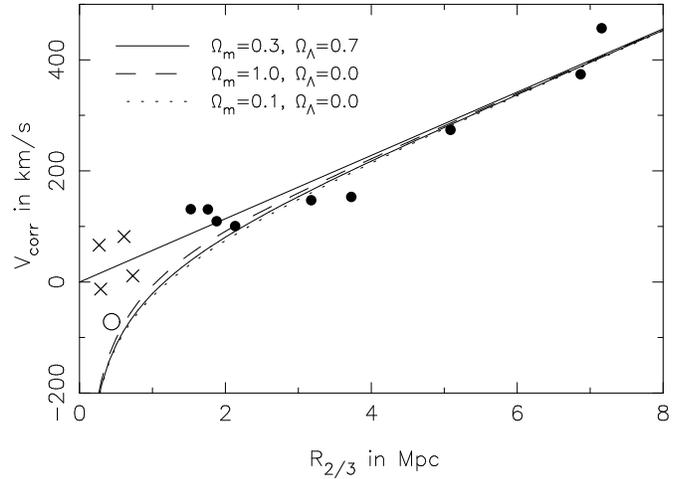}}
\caption{As Fig~\ref{F2}. Now we have added three theoretical
predictions. Each of them is based on a point mass model
assuming that the mass of the Local Group is
$M_\mathrm{LG}=2\times10^{12}M_{\sun}$. We examined three
cosmologies: 1) the Friedman model with $\Omega_m=0.3$
and $\Omega_\Lambda=0.7$ (the solid curve), 
2) the classical flat model with $\Omega_m=1.0$
and $\Omega_\Lambda=0.0$ (the dashed curve) and 
3) a low mass density model with
$\Omega_m=0.1$ and $\Omega_\Lambda=0.0$ (the dotted curve). 
}
\label{F3}
\end{figure}

Fig.~\ref{F3} shows the predictions for the velocity perturbations in
Sandage's point-mass model for the case when the mass of the Local Group
$M_\mathrm{LG} = 2 \times 10^{12} M_{\sun}$. For example,
van den Bergh (\cite{vdBergh99}) estimates that
$M_\mathrm{LG} = 2.3\pm0.6 \times 10^{12} M_{\sun}$.
Evans et al. (\cite{Evans00}) give an upper limit 
$M_\mathrm{LG} = 2.4 \times 10^{12} M_{\sun}$.
  
The velocities were deduced for three cosmologies using a numerical
algorithm for the Tolman-Bondi calculations 
(Hanski et al. \cite{Hanski00}). 
The solid curve shows the prediction for the
now preferred Friedman model 
$\Omega_m=0.3$ and $\Omega_{\Lambda}=0.7$. The dashed line corresponds to
the classical flat model $\Omega_m=1.0$ and $\Omega_{\Lambda}=0.0$.
The dotted line is the prediction for a low mass-density universe
$\Omega_m=0.1$ and $\Omega_{\Lambda}=0.0$. $\Omega_m$ is the mean
density parameter of the matter and $\Omega_\Lambda$ is
the density parameter induced by the $\Lambda$-term. One may note
that these models differ little from each other, which shows that the
value of the velocity deflection (and the zero-velocity distance) is
essentially determined by the mass of the LG.

Only galaxy that shows any significant deviation from the expected
Hubble law is \object{IC1613} but as pointed out in Sect.~2 its
dynamical status is not well known. On the other hand \object{M31}
deviates only little from the Hubble law (in fact it is within the
$1\sigma$ limit).  

%
%
%
%
%
%
%
 Sandage (\cite{Sandage86})
discussed the position of the barycentre. He pondered whether
it would be possible to reduce the scatter in the Hubble diagram
by ``changing the origin of the distances to the centroid as
well".  We have checked the behaviour of the velocity dispersion
$\sigma_v$ as a function of the position of the barycentre
between \object{Galaxy} and \object{M31}.

Recently Evans et al. (\cite{Evans00}) claimed that the halo
of \object{M31} is about as massive as the halo of
\object{Galaxy}. In this case the centroid would be
at the middle distance. Now, $\sigma_v$ changes quite slowly
in the region between $R_{1/2}$ and $R_{2/3}$. 
For example at $R_{1/2}$ 
the dispersion $\sigma_v=44\mathrm{\ km\,s^{-1}}$
and at $R_{1/4}$ 
it is $\sigma_v=49\mathrm{\ km\,s^{-1}}$. It is thus not
possible to make any statistically relevant claims about the
exact position of the barycentre.   
With a larger sample of
field galaxies, the position of the barycentre (and hence the
mass ratio 
$M_\mathrm{M31}/M_\mathrm{Galaxy}$) could perhaps be determined.

Finally we note that due to the lack of galaxies with
Cepheid distances within
$R_{2/3}=1\mathrm{\ Mpc}$ and $R_{2/3}=2\mathrm{\ Mpc}$
it is not possible to determine the zero-velocity surface.
It is though important to note that galaxies within 
$R_{2/3}=1\mathrm{\ Mpc}$ show practically no deflection as
compared to the theoretical curves shown in Fig.~\ref{F3},
except possibly the ``free-floating" \object{IC1613}, which could fit
some theoretical curve with $M_\mathrm{LG}<10^{12}M_{\sun}$.
\section{Discussion and Conclusions}

We may conclude that all the properties of the local Hubble flow, mentioned
by Sandage, become even more remarkable when more accurate distances
are used. Sandage (\cite{Sandage86}) 
noted that the derived velocity dispersion
around the local Hubble law has progressively decreased together
with increasing accuracy in distances, 
from Hubble's $200\mathrm{\ km\,s^{-1}}$ to
his $60\mathrm{\ km\, s^{-1}}$.  
Our results confirm that this trend continues, and
$\sigma_v \leq 40\mathrm{km\, s^{-1}}$.

The velocity-distance diagrams (Figs.~\ref{F2} and~\ref{F3})  
show that the local Hubble
constant is much the same as the more global value derived from the
Tully-Fisher indicator from the KLUN sample 
(Theureau et al. \cite{Theureau97};
Ekholm et al. \cite{Ekholm99b}).
This also confirms Sandage's (\cite{Sandage99}) recent estimate
that the very local Hubble constant is the same as the global $H_0$
within 10 percent.

The third important feature is the small distance where the Hubble
law emerges. The new data show that the linear Hubble law extends down
to at least 1.5 Mpc. Hence, within the standard Friedman model
(including the cosmological constant $\Lambda$; see Fig.~\ref{F3}), 
one gets
an upper limit for the mass of the Local Group ($\approx 2 \times
10^{12} M_{\sun}$).

The quiet Hubble flow within the very clumpy local galaxy universe
has always been a real riddle 
(Sandage et al. \cite{Sandage72}; Sandage \cite{Sandage99}). 
The problem of the local quiet Hubble flow was studied by
Governato et al. (\cite{Governato97}) using high-resolution
CDM N-body simulations.They constructed a large sample of
``Local Groups" and calculated the velocity dispersions in
$5\mathrm{\ Mpc}$ volumes around the LG candidates. They found
for $\Omega=1$ CDM model that the velocity dispersion is
$300-700\mathrm{\ km\,s^{-1}}$ and for $\Omega=0.3$ CDM model
$150-300\mathrm{\ km\,s^{-1}}$. They state that these simulations
were unable to produce a single LG having a velocity dispersion as
low as observed. 

The very local Hubble diagram offers several
important applications for cosmology when the number of accurate
Cepheid distances to local galaxies is increased.  The mass of the Local
Group and the position of its barycenter may be determined more
precisely.
And recently it has been suggested  (Chernin et al. \cite{Chernin00})
that the quietness of the local Hubble flow is a signature of the
cosmological vacuum (Einstein's $\Lambda$-constant corresponding to
energy density $\rho_{\Lambda} = \mathrm{const}$)
dominated universe
where the velocity perturbations are 
adiabatically decreasing. For solution of
both problems --the  small velocity dispersion and the Hubble law starting immediately at 
$1.5$ Mpc with the global $H_0$ -- 
Baryshev et al. (\cite{Baryshev00}) proposed that there is a
cosmological, homogeneous dark energy (quintessence) component 
with
time-variable energy density $\rho_Q(t)$ and equation of state
$p_Q = w~\rho_Q c^2 $ with $w \in [-1,0]$.
These discussions show that the very local
volume is extremely important for the study of global properties
of the universe.

%
%
%
%
\begin{acknowledgements}
This work was partly supported by the Academy of Finland
(project 45087: ``Galaxy Streams and Structures in the nearby
Universe" and project ``Cosmology in the Local Galaxy Universe").
We have made use of the Lyon-Meudon Extragalactic Database LEDA
and the Lyon Extragalactic Cepheid Database. We are grateful for the 
referee, M. Capaccioli, for his comments.
\end{acknowledgements}
%
%

%
%
\end{document}